\documentclass[12pt,a4paper,final]{article}
\usepackage[title,titletoc,toc]{appendix}
\usepackage{color}
\usepackage{hyperref} 
\usepackage[utf8]{inputenc}
\usepackage{amsmath}
\usepackage{amsfonts}
\usepackage{amssymb}
\usepackage{enumitem}

\usepackage{color}

\usepackage{tikz}
\usetikzlibrary{arrows,shapes}
\usetikzlibrary{trees}
\usetikzlibrary{matrix,arrows} 				
\usetikzlibrary{positioning}				
\usetikzlibrary{calc,through}				
\usetikzlibrary{decorations.pathreplacing}  
\usepackage{pgffor}							

\usetikzlibrary{decorations.pathmorphing}	
\usetikzlibrary{decorations.markings}
\tikzset{
    vector/.style={decorate, decoration={snake}, draw},
	provector/.style={decorate, decoration={snake,amplitude=2.5pt}, draw},
	antivector/.style={decorate, decoration={snake,amplitude=-2.5pt}, draw},
    fermion/.style={draw=black, postaction={decorate},
        decoration={markings,mark=at position .55 with {\arrow[draw=black]{>}}}},
    fermionbar/.style={draw=black, postaction={decorate},
        decoration={markings,mark=at position .55 with {\arrow[draw=black]{<}}}},
    fermionnoarrow/.style={draw=black},
    gluon/.style={decorate, draw=black,
        decoration={coil,amplitude=4pt, segment length=5pt}},
    scalar/.style={dashed,draw=black, postaction={decorate},
        decoration={markings,mark=at position .55 with {\arrow[draw=black]{>}}}},
    scalarbar/.style={dashed,draw=black, postaction={decorate},
        decoration={markings,mark=at position .55 with {\arrow[draw=black]{<}}}},
    scalarnoarrow/.style={dashed,draw=black},
    electron/.style={draw=black, postaction={decorate},
        decoration={markings,mark=at position .55 with {\arrow[draw=black]{>}}}},
	bigvector/.style={decorate, decoration={snake,amplitude=4pt}, draw},
}

\tikzstyle{block} = [draw, rectangle, 
    minimum height=3em, minimum width=6em]

\usepackage[font=small,labelfont=bf]{caption}

\hypersetup{ colorlinks=true, linkcolor=blue, citecolor=red }
\newcommand{\hf}{{\frac{1}{2}}}
\newcommand{\p}{\partial}
\newcommand{\be}{\begin{equation}}                                             
\newcommand{\br}{\begin{eqnarray}}                                             
\newcommand{\ee}{\end{equation}}                                               
\newcommand{\er}{\end{eqnarray}}

\begin{document}

\title{
\hfill\parbox{4cm}{\normalsize IMSC/2018/09/14}\\
\vspace{1.5cm}
Holographic Beta functions for the Generalized Sine Gordon Theory}
\author{{Prafulla Oak\footnote {prafullao@imsc.res.in} } and B. Sathiapalan{\footnote{bala@imsc.res.in}} \\ \\
  \small Institute of Mathematical Sciences\\  
  \small CIT Campus, Taramani, Chennai-600113\\
  \vspace{.5cm}  
  \small India \\
  \vspace{.5cm}
\small and\\
\small Homi Bhabha National Institute\\
\small Training School Complex, Anushaktinagar\\
\small Mumbai-400094,\\
\small lndia }
\date{}
\maketitle
\begin{abstract}
The Sine Gordon theory is generalized to include many scalar fields and several cosine terms. This is similar to the world sheet description of a string propagating in a tachyon background. This model  is studied as a (boundary) 2d euclidean field theory and also using an $AdS_3$ holographic bulk dual. The beta functions for the cosine vertex of this modified theory are  first computed in the boundary using techniques based on the exact RG. The beta functions  are also computed holographically using position space and  momentum space techniques. The results are in agreement with each other and with earlier computations. The cosine perturbation is of the form $\cos b X$. Due to wave function renormalisation the parameter $b$, and thus the dimension of the cosine, get renormalised. The beta function for this parameter is thus directly related to the anomalous dimension of the $X$ field. We compute this beta function in position space. They match with the earlier results in \cite{Oak:2017trw}.  
\end{abstract}

\tableofcontents 
\newpage

\section{Introduction}

The Sine-Gordon model and it's renormalization group evolution is interesting from many viewpoints. It is related to the X-Y model in two dimensions and gives the famous Kosterlitz-Thouless flow. A lot of work on both has been done before \cite{Amit:1979ab}-- \cite{Oak:2017trw}, \cite{polyakov}--\cite{ Dimock:1991ig}. It is also intimately related to string theory where a generalized version of the Sine-Gordon theory describes the bosonic string propagating in a tachyonic background. There also the RG flow is interesting because the $\beta$-function equations are related to the tachyon equation of motion.  

The form of the $\beta$- functions for the Sine-Gordon model are well known and have been computed by both field theoretic methods\cite{Amit:1979ab} long back and more recently using the Exact Renormalization Group (ERG)\cite{Oak:2017trw}. An interesting computation would be to reproduce them holographically in order to understand better the holographic RG.

It has been shown in \cite{Sathiapalan:2017frk} that  an ERG equation  in a boundary theory  can be mapped to a
scalar field action in AdS space time. The main results are for a free theory. Some suggestions for how the interactions should work out were given there. To understand these issues better it is important to understand RG equations in the boundary theory and obtain them from some bulk computations. The precise connection between these equations and what is called ``holographic RG" - which is really a radial evolution equation of the bulk theory - needs to be understood better. This paper is a
step towards that goal. There is extensive literature on the AdS/CFT correspondence and holographic RG \cite{Maldacena:1997re,Witten:1998qj}, \cite{Freedman:1998tz}--\cite{Arutyunov:1998ve}, \cite{Kiritsis:2016kog}--\cite{Bourdier:2013axa}.

The boundary theory is a free CFT perturbed by some composite (cosine) operators.  The bulk theory that reproduces the leading two and three point correlators is a cubic theory. Of course  there are any number of composite operators with definite scaling dimension and so the bulk theory should have a field of definite mass corresponding to each of these - we are assuming that an AdS dual exists for the free scalar theory in 2 dimensions.
One can study the RG flow of this theory and one should be able to reproduce the $\beta$ - function of the cosine operator of the boundary theory.  We do this calculation in this paper. 

However, motivated by the string theory tachyon connection we consider a generalized Sine-Gordon theory. In string theory,
instead of one scalar field, there are $D$ scalar fields ($D=26$ for the bosonic string). The tachyon perturbation is of the form
\[
\int d^2z ~ \int _k \phi(k) e^{i\vec k .\vec X }
\]
$\phi(k)$ is the tachyon field in momentum space. We can consider a continuum of values of $\vec k$. For each value of $k$ it corresponds to a Sine-Gordon like theory. In \cite{Das:1986cz} this theory was considered and shown to reproduce the
leading non linear terms in the tachyon - dilaton system equations of motion in string theory. 

Holographic techniques in position space are well suited for calculating correlation functions. In \cite{Sathiapalan:1987ic} a proper time method was used to evaluate the tachyon equation of motion starting from two point functions. We will use this technique here. For near marginal operators the two point function has the form
\[
\langle O_i(R) O_j(0) \rangle = \frac{G_{ij}}{R^4} + \frac{H_{ij}}{R^4} \ln \frac{R}{a}
\]
$G_{ij}$ is the Zamoldchikov metric. A similar formula exists for the open string boundary CFT, with $R^4$ replaced by $R^2$. In \cite{Sathiapalan:1987ic} it was shown (in the context of the open string) that 
\[ 
H_{ij}\phi^j=0
\]
 is the tachyon equation of motion to all orders in perturbation theory.  Furthermore, it was argued by Polyakov \cite{polyakov} (for closed strings) that the equation of motion and $\beta$-function are related simply:
\[
\frac{\p \Gamma[\phi]}{\p \phi ^i} = G_{ij}\beta ^j
\]
This was also shown to all orders in perturbation theory in the open string context in \cite{Sathiapalan:1987ic}. Thus we can conclude \cite{Sathiapalan:1995jh} that 
\[
H_{ij}\phi^j= G_{ij}\beta^j
\]
Thus to extract the beta function we can compute the two point function, corrected by interactions, and obtain the leading logarithmic deviation from the $\frac{1}{R^4}$ scaling to obtain the $\beta$ function. In the position space holographic calculation
we employ this technique.

Once the perturbation is turned on it is no longer a CFT. This should reflect itself in the bulk deviations from AdS. This requires taking into account the gravitational back reaction. This back reaction in the bulk can be seen to manifest itself in the field strength renormalization of the boundary scalar fields. This gives us the beta function for the field strength renormalization. To compute this we look at the fluctuations of the graviton about the AdS. This contribution comes from another cubic vertex in the bulk. This is also equivalent to the dilaton equation in the string theory context.

In section 2 we start with a brief overview of the Sine-Gordon model. We fix propagators and other normalizations. Then we give a brief summary of the calculation of the beta functions using ERG. Then we give the leading term of the beta function which has been computed in many ways in earlier papers as mentioned above. Next we motivate and detail the modification of the Sine-Gordon model. Then we compute the sub-leading term and extract the beta function. This concludes our boundary calculation.

In section 3 we give a brief overview of AdS-CFT computations using position space techniques. Then we compute the leading and sub-leading terms for the beta function from the bulk. In section 4 we start by briefly introducing computational techniques in AdS-CFT in momentum space. Then we calculate beta functions using momentum space techniques. All calculations are found to be in agreement. 

In section 5 we compute the beta function for the running of the field strength renormalization. This calculation is done in position space. This is found to be in agreement with previous results.

In  the appendix we determine relative normalizations between couplings in the bulk and boundary by comparing the generating functions for two-point functions computed from both sides. To fix the value of the coupling strength of the interaction vertex in the bulk we compare the generating function for the three point correlator on both sides. Doing this thus determines the bulk dual that reproduces correlations of the composite operators calculated in the free theory.

\pagebreak

\section{The ERG beta function calculation of the generalized Sine-Gordon model.}

\subsection{The generalized Sine-Gordon model}

The action for the generalized theory is

\begin{align}
S_{boundary}=&   \frac{1}{4\pi}    \int d^2 x    [   (\partial_{\mu} \vec{X}) .  (\partial^{\mu} \vec{X})  + m^2 \vec X .\vec X  +   \frac{F}{a(0)^2} \cos ~ (\vec{b}_1 . \vec{X} )     \\                       \nonumber
& +   \frac{G}{a(0)^2} \cos ~ (\vec{b}_2 . \vec{X} )  +   \frac{H}{a(0)^2}       \cos ~ (\vec{b}_3 . \vec{X} )           ]  
\end{align}

in euclidean d=2. Powers of $a(0)$, the UV cutoff, have been added so that the engineering dimension of the action is zero. The mass term acts like an IR regulator in the propagator. In our calculations we cut off all integrals in the IR by a moving scale, therefore we encounter no IR divergences. At marginality all $b_i^2=4$. This can be viewed as a world sheet action for a string in the presence of a background tachyon field with some definite momentum \cite{Das:1986cz}. The marginality condition is the ``on-shell" condition for the tachyon. In that case the metric, $g_{MN}$, for the dot product of $b^M b^N g_{MN}$ has Minkowski signature.  By doing this we have an additional freedom to tune the norm of the vector to the required value by modifying the individual components of the vector. This is important for the massless and higher string modes though it is not required for the tachyon. From here on all $b_i$'s and $X$'s  are understood to be vectors - $b_i^M,X^M$. We will drop all arrows on the top and suppress the vector index.

 We want to calculate beta functions for the flow of $F$ and $b_1$. Due to wave function renormalisation the parameter $b_1$, and thus the dimension of the cosine, get renormalised. The beta function for this parameter is thus directly related to the anomalous dimension of the $X$ field. F gets corrections from the self interaction of the cosine and corrections from higher order terms. From the string point of view, we are computing scattering amplitudes for the zeroeth mode of the closed strings with momenta $b_i$ at position $X(x)$, where $\exp i b_i  .   X(x)$ is the vertex operator for the tachyon. $\exp i b_iX(x)$ is a tachyon vertex operator for a distinct closed string, each with momentum $b_i$, where now instead of b being continuous, as in the introduction, we choose a discrete set of $b_i$'s. We will choose $b_1+b_2+b_3=0$ for reasons that will become clear.
 It has been shown in \cite{Amit:1979ab} that the Sine-Gordon theory is renormalizable with a  well defined expansion in $F$ and a parameter $\delta =\frac{ b^2}{4}-1$.

We want to reproduce the beta functions of the Sine-Gordon model from the bulk. In the boundary theory we will look at the action of the generalized Sine-Gordon theory and use that to compute the beta function. In the $\cos b.X$ term, the dot product is over an N dimensional vector space and as such the b's and the X's all are vectors under some Lie group, such as SO(N). With $N$ scalar fields, the central charge, $c$,  of the free CFT is $N$.  In $AdS_3$ a large $c$-expansion plays the role of large $N$ in the more familiar $AdS_5$ case. We could do an expansion of the boundary theory in this N. Thus we can invoke the AdS-CFT correspondence to, as explained in the introduction, compute the appropriate multi-point boundary correlators from the bulk, to get the beta functions. These correlators diverge when some or all of these points coincide. Thus, to extract the beta function we compute the leading logarithmic deviation from the $\frac{1}{R^4}$ scaling of these correlators. In the position space holographic calculation we employ this technique.

When $b_1+b_2+b_3=0$, the first non vanishing higher point correlator is the cubic one involving all three cosines. From the point of view of the string theory tachyon, this constraint on the $b_i$ is just momentum conservation. From the CFT viewpoint, this comes from integrating over the zero mode of $X(x)$. In that case the beta function for $F$ starts at quadratic order
\[
\beta_F \approx O(GH)
\]
This is the cubic term in the tachyon equation of motion \cite{Das:1986cz}.
At higher orders the four point correlator is always non zero and there is a contribution of $O(F^3)$. (This is the first sub-leading term in the usual Sine-Gordon model.)

\subsection{The propagator, other preliminaries.}

We start with the kinetic term

\begin{equation}
S_{Kinetic}= {1\over 2\alpha'}\int d^2x ~\partial_{\mu} X \partial^{\mu}X
\end{equation}
$\alpha'$ is like the string tension. The propagator is

\begin{equation}
G^{MN}(x_1,x_2)=< X^M(x_1) X^N(x_2)> = -g^{MN}{\alpha'\over 2\pi} \ln {|\vec{x}_1-\vec{x}_2| \over    L    }
\end{equation}

Set $\alpha'=2\pi$. L is an arbitrary scale to make the argument of log dimensionless.

Therefore,

\begin{equation}
\left\langle  : \cos b.X(x_1):     : \cos b.X(x_2)  :    \right\rangle  =  \hf   \left(   \frac{|\vec{x}_1 -  \vec{x}_2 |}{L}        \right)         ^{-b^2}  
\end{equation}

The mass dimension of a marginal operator in d=2 is 2. Therefore, $b^2/2=2$.   The beta functions are a power series expansion in the two couplings $F$ and $\vec{b}$. F is a small number, $F \to 0$. $b^2=4$ is a large number, therefore we will look for a suitable expansion parameter which is small. Both parameters are renormalized. When $F$ is non zero, the theory is interacting and there is a wave function renormalization that causes $\delta$ also to run.

\subsection{The ERG calculation.}

The ERG can be described by

\begin{equation}
\psi(X,t)=e^{-\hf   \int d^2 x_1 d^2 x_2     F_{x_1x_2t}       \frac{\delta}{\delta X(x_1)}      \frac{\delta}{\delta X(x_2)}                 } \psi(X,0)
\end{equation}

Here

$F_{x_1x_2t}=-\hf ln\frac{(x_1-x_2)^2+a(t)^2}{(x_1-x_2)^2+a(0)^2}$ is the ERG high energy "propagator". $t$ is the scale upto which you are doing the RG transformations, $a(0)$ is the UV cutoff, $a(t)$ is the IR cutoff. $a(t)=a(0) e^{t}$. Which implies $t=ln(a(t)/a(0)$, the log of the ratio of the scales whose coefficient is the beta function.

\begin{align}
\psi(X,0)=  e^{  -  \int d^2 x     \frac{1}{4\pi}    [   \frac{F}{a(0)^2} \cos ~ (\vec{b}_1 . \vec{X} )      +   \frac{G}{a(0)^2} \cos ~ (\vec{b}_2 . \vec{X} )  +   \frac{H}{a(0)^2}       \cos ~ (\vec{b}_3 . \vec{X} )        ]                }         
\end{align}

is the un-integrated "partition function" of the theory and the evolution operator $e^{-\hf   \int d^2 x_1 d^2 x_2     F_{x_1x_2t}       \frac{\delta}{\delta X_1}      \frac{\delta}{\delta X_2}                 }$ acting on $\psi(X,0)$ upto some scale $t$, gives $\psi(X,t)$, thus implementing the RG.

One can bring down appropriate powers of cosine from the exponential and act on it with the ERG operator. The calculation can then be organised as the ERG operator acting on a power series

\begin{align}
&   e^{-\hf   \int d^2 x_1 d^2 x_2     F_{x_1x_2t}       \frac{\delta}{\delta X(x_1)}      \frac{\delta}{\delta X(x_2)}                 }     [          \int   \frac{d^2 x_1}{a(0)^2}            (a1) \cos bX(x_1)         \\         \nonumber
& +        \int         \frac{d^2 x_1}{a(0)^2}        \frac{d^2 x_2}{a(0)^2}        (a2)  \cos bX(x_1) \cos bX(x_2)                                   \\         \nonumber
&        +                \int                \frac{d^2 x_1}{a(0)^2}        \frac{d^2 x_2}{a(0)^2}            \frac{d^2 x_3}{a(0)^2}                   (a3) \cos bX(x_1)  \cos bX(x_2)  \cos bX(x_3)]   
\end{align}

 the $ai$'s, most generally, being the different corresponding coefficients. We look for corrections to cos bX which is a term of the form

\begin{equation}
(c1+~c2+~c3) t       \int     \frac{d^2 x_1}{a(t)^2}            \cos bX(x_1)
\end{equation}

where c1, c2 and c3 are the coefficients obtained after the ERG operator acts on the power series term by term. The final expression can be reorganized such that $(c1+c2+c3) ~t$ is the correction to the coupling(the above equation) and its derivative w.r.t to t is the beta function. Further details can be found in \cite{Oak:2017trw} and \cite{Igarashi:2009tj}.

For our case, to calculate the leading contribution we bring down one power of $F cos b_1X$ and apply the ERG operator to it.

\subsection{Leading term in $\beta_{F}$.}

The ERG operator acting on the interaction term gives

\[
\int {d^2x_1\over a(0)^2}   {F \over 4\pi}        \exp(-\hf     \int d^2 x_i d^2 x_j       F_{x_ix_jt}\frac{\delta^2}{\delta X(x_i) \delta X(x_j)})      cos({bX(x_1)})
\]

Simplifying we get,

\begin{equation}
\int {d^2x_1\over a(t)^2}   {F \over 4\pi}                   (1-2\delta t)      cos({bX(x_1)})
\end{equation}

The leading term in the $\beta$-function for F is,
\be	\label{betaalpha1}
\beta_{F} =-2\delta F
\ee

$\delta= {   b^2 \over 4}-1$ is the other small expansion parameter in terms of which we will calculate beta functions. We refer the reader to \cite{Oak:2017trw} for further details.

\subsection{The sub-leading term.}

To calculate the sub-leading contribution we will bring down one power of $     \frac{G}{a(0)^2} \cos~b _2. X (x) $  and   $   \frac{H}{a(0)^2}  \cos~   b_3 . X (x) $ each.

\begin{align}\label{evolu oper on two cosines}
    \frac{GH}{(4 \pi)^2}        \int    \frac{d^2x_1 d^2x_2}{a(0)^4}     \exp           \left(            -\hf      \int d^2 x_i d^2 x_j          F_{x_ix_jt}\frac{\delta^2}{\delta X(x_i) \delta X(x_j)}             \right)           \cos b_2 X(x_1)  ~~\cos b_3 X(x_2)
\end{align}

\begin{align}
=   \frac{GH}{4}      t     \frac{1}{(4 \pi)}        \int   \frac{d^2x }{a(t)^2}  \cos b_1 X(x)
\end{align}

Refer to Appendix \eqref{app ERG sub leading beta F} for further details.

\subsection{The beta function.}

We can organize the calculation as follows,

\begin{equation}
   \left(      F       (1-2\delta t)      +     \frac{GH}{4} t               \right)
         \frac{1}{4\pi}                            \int {d^2x_1\over a(t)^2}     cos({bX(x_1)})
\end{equation}

Therefore, the full beta function, the t-derivative of the coefficient of the above expression, is,

\begin{equation}\label{betaFERG}
\beta_F=      -    (2F\delta       -     \frac{GH}{4}   )
\end{equation}

\section{Position space calculation of the beta function from the bulk.}

\subsection{A brief overview of AdS/CFT and the beta function calculation.}

The AdS metric in the Poincare patch is

\begin{equation}\label{Poincare}
ds^2=\frac{1}{z^2} [dz^2+d\vec{x}^2]
\end{equation}

$\vec{x}$ is euclidean.

According to the AdS-CFT correspondence

\begin{equation}
\int  D\Phi \exp(-S[\Phi])= \left\langle   \exp \left(  -\int  _{\partial AdS}  \phi_0    O                  \right)       \right\rangle   
\end{equation}

which to leading order is

\begin{equation}
S_{bulk}[\phi_0]=- W_{QFT}[\phi_0]
\end{equation}

where $S_{bulk}[\phi_0]$ is the bulk action and $W_{QFT}[\phi_0]$ is the connected generating functional of the boundary theory.

The correlation funtions from the bulk are calculated by taking variations w.r.t $\phi_0$ on both sides. A general n point correlation function is given by
\begin{equation}
\left\langle O_1(x_1)...O_n(x_n) \right\rangle  =   (-1)^{n+1} \frac{\delta^n S_{bulk}}{\delta \phi_0(x_1)...\delta \phi_0(x_n)}       |_{\phi_0=0}
\end{equation}

To compute beta functions from the bulk we start with the bulk action with the $ \phi \chi \gamma$ term,

\begin{align}
S_{bulk}=& \int d^3 x       \sqrt{g}      [     \hf  (\partial \phi)^2 + \hf (m_{\phi}\phi)^2        + \hf  (\partial \chi)^2 + \hf (m_{\chi}\chi)^2              \\                     \nonumber                                     
&+ \hf  (\partial \gamma)^2 + \hf (m_{\gamma}\gamma)^2 -        \lambda_3 \phi \gamma \chi]
\end{align}

The $\phi$, $\gamma$ and $\chi$ correspond to $\cos b_1X$, $\cos b_2X$, $\cos b_3X$ respectively.

The free equation of motion is

\begin{equation} \label{bulkeom1}
\frac{1}{\sqrt{g}}\partial_\mu \big ( \sqrt{g} g^{\mu \nu} \partial_\nu \phi)- m^2 \phi=0
\end{equation}

with boundary conditions

$\phi(z_0,\vec{z})=0$ for $z\rightarrow \infty $ and $\phi(z_0,\vec{z})\rightarrow z_0^{d-\Delta} \phi_0(\vec{z})$ as $z_0 \rightarrow 0$

The normalized bulk to boundary Green's function is

\begin{equation} \label{sclrprop}
K_\Delta(z_0,\vec{z},\vec{x})= \frac{\Gamma(\Delta)}{\pi^{d/2} \Gamma(\Delta-d/2)}  \bigg (  \frac{z_0}{z_0^2+(\vec{z}-\vec{x})^2}  \bigg )^\Delta
\end{equation}

The solution to \eqref{bulkeom1} is

\begin{equation}
\phi(z_0,\vec{z})= \frac{\Gamma(\Delta)}{\pi^{d/2} \Gamma(\Delta-d/2)}   \int d^2x   \bigg (  \frac{z_0}{z_0^2+(\vec{z}-\vec{x})^2}  \bigg )^\Delta \phi_0(\vec{x})
\end{equation}

To extract the beta function we first regulate the generating function for the correlators of the boundary theory to be calculated from the bulk by inserting $x_0$ which acts as the UV cutoff. Then we compute the generating function for the two point function with one particle offshell and then, as was described in the introduction, obtain the leading logarithmic deviation from the $\frac{1}{R^4}$ scaling which comes, at the leading order, from taking the particle offshell. Then we extract terms which are logarithmically divergent in terms of $x_0$ which acts as the UV scale.

\begin{center}

\begin{tikzpicture}[scale=2.5]

\draw [fermion] (-1,0) -- (0,0);
\node at (-.5 ,.2) {$\vec{b}$};
\node at (-.9,0.1) {$\vec{x}_1$};

\draw [fermion] (1,0) -- (0,0);
\node at (.5 ,.2) {$\vec{b}$};
\node at (.9,0.1) {$\vec{x}_2$};

\draw  (0,0) circle (1);

\node at (0,-1.4) { Fig. 1. } ;
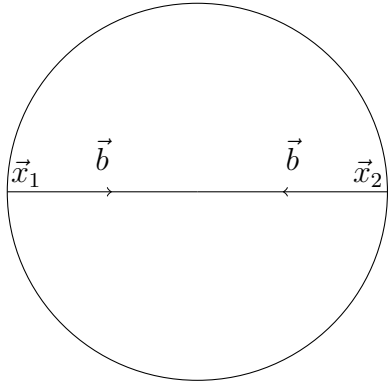
\node at (0,-1.6) {The leading order Witten diagram};

\draw [fermionbar] (3 ,0) -- (4,0);
\node at (3.9,0.1) {$\vec{x}_1$};
\node at (3.5 ,.2) {$\vec{b}_1$};

\draw  (3,0) circle (1);

\draw [fermionbar](3,0) node[below] {$\lambda_3$}  -- (2.35,.8) ;
\node at (2.4,0.6) {$\vec{x}_2$};
\node at (2.85 ,.4) {$\vec{b}_2$};

\draw [fermionbar](3,0) -- (2.35,-.8);
\node at (2.6,-0.8) {$\vec{x}_3$};
\node at (2.8 ,-.45) {$\vec{b}_3$};

\node at (3,-1.4) { Fig. 2. } ;
\node at (3,-1.6) {Witten diagram for }      ;
\node at (3,-1.8) {the sub-leading contribution};

\end{tikzpicture}

\end{center}

\subsection{Leading order}

The bulk action for the free massive scalar is

\begin{equation}
S=\int         d^{d+1}y \sqrt{g} \bigg [   \big (\frac{1}{2} g^{\mu \nu} \partial_\mu \phi(y) \partial_\nu \phi(y) +\frac{1}{2}m^2 \phi^2(y)        \big )    \bigg]
\end{equation}

Evaluating the free action on-shell we get,

\begin{align}
&\hf    \phi_0^2    \int            d^dx_1 d^dx_3                d^d y dy_0     \bigg  [           \partial_{\mu}     \left(        y_0^{-d+1} K_{\Delta}(y_0,\vec{y};\vec{x_1})    \partial^{\mu} K_{\Delta}(y_0,\vec{y};\vec{x_3})     \right)         \bigg     ]
\end{align}

 Choose the outward pointing normal along the radial direction and  by the Gauss's divergence theorem, do the surface integral.

\begin{align}
&S=-\hf    \phi_0^2    \int            d^dx_1 d^dx_3                d^d y    \bigg  [                y_0^{-d+1} K_{\Delta}(y_0,\vec{y};\vec{x_1})   \partial _0       K_{\Delta}(y_0,\vec{y};\vec{x_3})          \bigg     ]   \bigg  |_{y_0=\epsilon}  
\end{align}

where $\epsilon \to 0$, therefore $y_0$ is close to the boundary, $y_0 \to 0$. We identify $y_0$ with $x_0$, the UV regulator. The minus sign comes from choosing the convention for the outward pointing normal $n^{\mu}=(-\epsilon, \mathbf{0}  )$. Therefore, using

\begin{equation}
\lim_{x_0 \to 0} x_0^{\Delta-d} K_{\Delta}(x_0,\vec{y};\vec{x}_i)      \to \delta^{(d)}(\vec{y}-\vec{x}_i)
\end{equation}

\begin{align}
&S=-\hf    \phi_0^2  C_{\Delta}  \int            d^dx_1 d^dx_3                d^d y                                      \bigg  [        \frac{\Delta  \delta^{(d)} (\vec{y}-\vec{x_1}) }{(x_0^2+(\vec{y}-\vec{x}_3))^{\Delta}}            -   \frac{(2\Delta ) x_0^{2}  \delta^{(d)} (\vec{y}-\vec{x_1})}{(x_0^2+(\vec{y}-\vec{x}_3))^{\Delta+1}}                    \bigg     ]          
\end{align}

This is the action for the free massive term. A general n-point correlation function is given by

\begin{equation}
\left\langle O_1(x_1)...O_n(x_n) \right\rangle  =   (-1)^{n+1} \frac{\delta^n S_{bulk}}{\delta \phi_0(x_1)...\delta \phi_0(x_n)}       |_{\phi_0=0}
\end{equation}

Therefore the generating function of a two point function will have another minus. The two point functions from the bulk and the boundary now match. The generating function for the two point function is,

\begin{align}
&S_2=-   ( -1)   \hf    \phi_0^2  C_{\Delta}  \int            d^dx_1 d^dx_3                                               \bigg  [        \frac{\Delta   }{(x_0^2+(\vec{x}_1-\vec{x}_3))^{\Delta}}            -   \frac{(2\Delta ) x_0^{2}  }{(x_0^2+(\vec{x}_1-\vec{x}_3))^{\Delta+1}}                    \bigg     ]          
\end{align}

The log divergent term comes from the first term and is retained. The second term is $x_0^2$ suppressed. In the limit $x_0 \to 0$, it vanishes. We drop this term.  $C_{\Delta_i}=\frac{\Gamma(\Delta_i)}{\pi^{d/2} \Gamma(\Delta_i-d/2)}$, therefore

\begin{equation}
S_2= \frac{\Gamma(\Delta+1)}{\pi^{d/2} \Gamma(\Delta-d/2)}   \int d^dx_1 d^d x_3    \frac{1}{(x_0^2+(\vec{x_1}-\vec{x_3})^2)^{\Delta}}               \frac{\phi_0^2}{2!}   
\end{equation}

This is the generating functional for the two point function with the UV regulator $x_0$. Setting $d=2$, $\Delta=2(1+\delta)$, $x_1 $ to zero and $ x_3$ to R, $R \to \infty$, the expression becomes,

\begin{align}
  \frac{\phi_0^2}{2!} \pi \Delta   (\Delta-1)  \int  d   \left( {  x_1^2     \over     x_0^2}    \right)         d    \left(    {x_3^2       \over x_0^2}    \right)     \frac{1}{    (  (    x_0^2    +R^2)/ x_0^2)^{2(1+\delta)}}
\end{align}

Multiply and divide by R and expand the denominator for $\delta \ll 1$ such that $(R/x_0)^{-4\delta}=1-4\delta \log R/x_0$, we get

\begin{align}
  \phi_0^2 \pi   \int            d   \left( {  x_1^2     \over     R^2}    \right)         d    \left(    {x_3^2       \over R^2}    \right)                                  \left(   1-4\delta \log\frac{R}{x_0}                       \right)
\end{align}

We will extract the leading term of the beta function from this.

\subsection{Order $\phi \gamma   \chi$}

The beta function is the change in the couplings of the theory under scaling transformations. To determine the deviation from the canonical scaling dimension we look at the behaviour of the two point function slightly away from marginality and determine the leading term of the beta function. This was the calculation we did above. To calculate the sub-leading term we start with a two point function, insert another operator, therefore now we have a three point function, and look for the log deviation from $1/R^4$ scaling for this object.
We first calculate  the generating function for the three point function. To do this we start with

\begin{equation}
S_3=       -     \lambda_3                  \int d^{d+1} y \sqrt{g} ~         \phi(y)            ~  \gamma(y)   ~  \chi(y)   
\end{equation}

where

\begin{equation}
\phi(z)=\phi(z_0,\vec{z})=   \int d^dx              K_{\Delta}(z_0,\vec{z},\vec{x})           \phi_0(\vec{x})
\end{equation}

and substitute this for $\phi(y)$, $\gamma(y)$, $\chi(y)$ in $S_3$. Here $\sqrt{g}=y_0^{-(d+1)}$, $d=2$, $C_{\Delta_i}=\frac{\Gamma(\Delta_i)}{\pi^{d/2} \Gamma(\Delta_i-d/2)}$. $\phi_0$, $\gamma_0$ and $\chi_0$ are the couplings of the boundary theory (boundary values of the bulk fields, they have no coordinate dependence). This becomes(details in Appendix \eqref{app sub leading beta f}),

\begin{align}
S_3=-         \lambda_3 \pi       \phi_0 \gamma_0   \chi_0              \int      d   \left( {  x_1^2     \over     R^2}    \right)         d    \left(    {x_3^2       \over R^2}    \right)                    \log \frac{R}{x_0}    
\end{align}

\subsection{The Beta function}

The full generating function can be organized as

\begin{equation}
S=S_2+S_3
\end{equation}

\begin{align}
=              ~    \pi  \int            d   x _1^2            d x_3^2                      \left(                \phi_0^2    -    2     \phi_0        (2 \phi_0  \delta                    +         {         \lambda_3  \gamma_0   \chi_0            \over        2        }                )                 \log \frac{R}{x_0}                          \right)    
\end{align}

Substituting the relations between $\phi_0$, $\gamma_0$, $\chi_0$ and F,G,H(\eqref{phi0}, \eqref{gamma0}, \eqref{chi0}) and the value of $\lambda_3$\eqref{lambda3} we get,

\begin{align}
=           {       1    \over                 64           }         \int            d   x _1^2            d x_3^2                      \left(             F^2       -    2     F       (2         \delta        F                 -       {        GH        \over                   4     }                )                \log \frac{R}{x_0}                       \right)    
\end{align}

We have calculated the correction to $F^2$. To get the beta function we want to isolate the change in $F$. To do this we note,

\begin{equation}
F^2   \to   F'^2   =F^2 +    \delta(F^2)             =   F^2        +2 F \delta(F)                   
\end{equation}

\begin{equation}
   F^2        +           (     -     4 \delta   F^2   +      F GH/2      )      \log      \frac{R}{x_0}          =F^2     -   2        F            (        2 \delta   F                 -            GH/4                      )                \log \frac{R}{x_0}              
\end{equation}

 and
comparing the above two expressions we see

\begin{equation}\label{deltaFforBetaF}
     \delta(F) =       - \left(   2 \delta  F       -      {    GH      \over      4}        \right)                   \log \frac{R}{x_0}   
\end{equation}

In the Poincare patch \eqref{Poincare} a physical distance s between two points is $s=s_{coord}/z_0$, where $s_{coord}$ is the coordinate distance between the points at the boundary at $z=z_0$. $x_0$ is the physical UV cutoff scale on the boundary where the metric is $\delta_{\mu \nu}$. On the boundary at $z=z_0$, the coordinate distance becomes $R=x_0 z_0$. If one identifies this with $a(t)=a(0)e^t$, the moving IR scale of the boundary theory, and one further identifies $x_0$ with a(0), then one can naturally identify the boundary position $z_0$ with $e^{t}$ and then moving along the z direction is the same as scaling transformations of the boundary theory, thus implying that moving along the z direction is the same as doing RG transformations. Therefore $\log \frac{R}{x_0}$ gets identified with t. Then the t-derivative of \eqref{deltaFforBetaF} gives us the beta function

\begin{equation}
     \beta_F =       - \left(   2 \delta  F       -      {    GH      \over      4}        \right)                  
\end{equation}

which matches our result from the boundary calculation \eqref{betaFERG}.

\section{Beta function computation using momentum space techniques from the bulk. }

We will now compute the results obtained in earlier sections using momentum space techniques. This method is more closely related to the conventional way of obtaining beta functions, namely, picking out the log divergence, introducing a renormalization scale and extracting the beta function from the derivative w.r.t the scale. Although using the previous approach, one can do computations for a larger class of bulk vertices, doing the calculation by this method is much simpler in some of the cases of interest. We will start with a brief introduction to AdS-CFT computations in momentum space.

\subsection{A brief summary of AdS/CFT from the momentum space perspective.}

The bulk action with the term $ \Phi \chi \gamma$ is (here we relabel the field $\phi$ from earlier to $\Phi$),

\begin{equation}
S_{bulk}= \int d^3 x         \sqrt{g}       [     \hf  (\partial \Phi)^2 + \hf (m_{\Phi}\Phi)^2        + \hf  (\partial \chi)^2 + \hf (m_{\chi}\chi)^2          + \hf  (\partial \gamma)^2 + \hf (m_{\gamma}\gamma)^2 -   \lambda_3 \Phi \gamma \chi]
\end{equation}

The equation of motion is
\begin{equation}
(-\Box_G+m^2)\Phi=\lambda_3\gamma \chi
\end{equation}

$\Phi$ can be expanded in powers of $\lambda$

\begin{equation}
\Phi=\Phi_{0}+\lambda_3 ~\Phi_1+...
\end{equation}

\begin{equation}
\Phi_{0}=\phi_{00} + z^2 \phi_{02}+ z^4 \phi_{04}+...
\end{equation}

\begin{equation}
\Phi_{1}=\phi_{10} + z^2 \phi_{12}+ z^4 \phi_{14}+...
\end{equation}

$\gamma$ and $\chi$ have similar expansions.

The equations of motion can be  solved perturbatively order by order in $\lambda_3$.
			
\begin{equation}
(-\Box_G+m^2)\Phi_0=0
\end{equation}

and

\begin{equation}
(-\Box_G+m^2)\Phi_1=\gamma \chi
\end{equation}

We fourier transform along all directions parallel to the boundary at $z=0$. We write the Fourier transform of $\Phi(z,\vec{x})$ as $\Phi(z,\vec{p})$. The free equation of motion becomes

\begin{equation}
L_{d,\Delta}(z,p)\Phi(z,p)=0
\end{equation}

where

\begin{equation}
L_{d,\Delta}(z,p)=-z^2 \partial_z^2+(d-1)z\partial_z+m^2+z^2p^2
\end{equation}

The bulk to boundary propagator is given by

\begin{equation}
\mathcal{K}_{d,\Delta}(z,p)=\frac{2^{\frac{d}{2}-\Delta+1}}{\Gamma(\Delta-\frac{d}{2})}p^{\Delta-\frac{d}{2}} z^{d/2} K_{\Delta-\frac{d}{2}}(pz)
\end{equation}

The bulk to bulk propagator is

\begin{equation}
\mathcal{G}_{d,\Delta}(z,p;\zeta)=(z\zeta)^{d/2}  I_{\Delta-d/2}(pz) K_{\Delta-d/2}(p\zeta)
\end{equation}

for $z\leq \zeta$

and

\begin{equation}
\mathcal{G}_{d,\Delta}(z,p;\zeta)=(z\zeta)^{d/2}  I_{\Delta-d/2}(p\zeta) K_{\Delta-d/2}(pz)
\end{equation}

for $z\geq \zeta$

We saw earlier that for small $\lambda_3$, $\Phi$ can be expanded in powers of $\lambda_3$

\begin{equation}
\Phi=\Phi_{0}+\lambda_3 ~\Phi_1+...
\end{equation}

$\Phi$ has a near boundary(small z) expansion

\begin{align}
\Phi=(\phi_{00} + \lambda_3 \phi_{10}   + O(\lambda_3^3)      ) +z^2 (\phi_{02} + \lambda_3 \phi_{12}    + O(\lambda_3^3)   ) + O(z^3) 
\end{align}

Therefore,

\begin{equation}
\phi_0=\phi_{00}+\lambda_3 \phi_{10}  +\lambda_3^2 \phi_{20} +...
\end{equation}

should be considered the full source, each of the $\phi_{i0}$ sourcing $\Phi_i$. The solutions to the equations of motion order by order in $\lambda_3$ are,

\begin{equation}
\Phi_0(z,\vec{p})=  \mathcal{K}_{d,\Delta}(z,p) \phi_0
\end{equation}

 To begin with we turn off the higher order terms in the source $\phi_0$. We only keep the leading term $\phi_{00}$ in $\Phi_{0}$ here. Later we will see that the higher orders will have to be turned on.

The solution to the second equation is

\begin{align}
\Phi_1= &  \int   \frac{d^dk_1   d^dk_2 }{(2 \pi)^{2d}}  \gamma_0                         \chi_0                                                                                                          \delta^{(d)}(k_1+k_2+k_3)          \\                \nonumber
&                   \int^{\infty}_{0}   \frac{d\zeta}{\zeta^{d+1}}  \mathcal{G}_{d,\Delta}(z,k_1;\zeta) \mathcal{K}_{d,\Delta}(\zeta,k_2)  \mathcal{K}_{d,\Delta}(k_3,\zeta)
\end{align}

Further details can be found in \cite{Bzowski:2015pba}.

\subsection{Leading order}

For $d=2$ and $\Delta=2+2\delta$, the solution to the free equation of motion is

\begin{equation}
\Phi_0=p^{1+2\delta}  z  K_{1+2\delta}(pz)      \phi
\end{equation}

We expand about $z=0$ and then expand for $\delta \ll 1$, we get

\begin{equation}
\Phi_0 =         (1-2\delta \log z)  \phi_0
\end{equation}

At $z=x_0$ where $x_0 \to 0$, $-2\delta   \phi_0   \log x_0   $ is divergent. Therefore $\phi_0$ would have to change to

\begin{equation}
  \phi_0      \to     \phi_0     +    2\delta      \phi_0         \log \frac{x_0}{R}       
\end{equation}

to cancel the divergent term. Here R is an IR scale introduced on dimensional grounds. Thus we see that a change in the canonical scaling dimension of $\Phi_0$, induces a flow in $\phi_0$.

\subsection{Order  $\phi \gamma  \chi$}

To get the contribution at this order we look at the solution of $\Phi_1$.

We want to solve the integral

\begin{equation}
I^{\delta, <}_{d=2,\Delta=2}=\int_{x_0}^z       \frac{d\zeta}{\zeta^{d+1}}                            \mathcal{G}_{d,\Delta}(z,p_1;\zeta) \mathcal{K}_{d,\Delta}(\zeta,p_2)  \mathcal{K}_{d,\Delta}(p_3,\zeta)
\end{equation}

in the near boundary region $\zeta \leq z$. $x_0$ is the UV cut-off.

\begin{align}
\Phi_{1}&= \gamma_0   \chi_0   \int_{x_0}^z       \frac{d\zeta}{\zeta^{d+1}}                            (z \zeta)^{d/2}  I_{\Delta_1 -d/2}(p_1 \zeta)     K_{\Delta_1 -d/2}(p_1 z)            \\                 \nonumber
&    \frac{2^{d/2-\Delta_2+1}}{\Gamma(\Delta_2-d/2)}      p_2^{\Delta_2-d/2}   \zeta^{d/2}                  K_{\Delta_2 -d/2}(p_2 \zeta)                                                                    \frac{2^{d/2-\Delta_3+1}}{\Gamma(\Delta_3-d/2)}      p_3^{\Delta_3-d/2}   \zeta^{d/2}                        K_{\Delta_3 -d/2}(p_3 \zeta)            
\end{align}

Since the $\delta$ has no effect at this order all $\Delta$'s are set to 2. The log divergent part of $\Phi_{1}$ is,

\begin{align}
    \hf        \gamma_0   \chi_0       p_1 z  K_1(p_1 z)          (  -   \log   x_0      )        
\end{align}

We can expand $\Phi$ in powers of $\lambda_3$. Therefore we can write the log divergent terms of $\Phi$,

\begin{equation}
\Phi=\Phi_{0}+\lambda_3     \hf        \gamma_0   \chi_0       p_1 z  K_1(p_1 z)          (  -   \log   x_0      )  
\end{equation}

This diverges as $x_0 \to 0$. $   p_1 z  K_1(p_1 z)    $ is the solution to the leading order equation of motion $\Phi_0$, therefore to make the full $\Phi$ finite we can turn on a subleading $O(\lambda_3)$ term in the expansion of the source $\phi_0$ in $\Phi_0$ (as mentioned before we are turning on subleading coefficients).

\begin{equation}
\phi_0=\phi_{00} + \lambda_3 \phi_{10} +...
\end{equation}

The modified source $\phi_0$ is

\begin{equation}
\phi_{0}=                (1+2\delta\log  x_0    )           \phi_{00}             +\lambda_3 \phi_{10}
\end{equation}

set 

\begin{equation}
 \phi_{10} =   \hf        \gamma_0   \chi_0                (     \log   x_0    ) 
\end{equation}

The modified source becomes

\begin{equation}
\phi_0=               \left(        1      +     2\delta           \log \frac{x_0}{R}              \right)             \phi_0                  +       \lambda_3       \hf          \gamma_0   \chi_0            (        \log \frac{x_0}{R}     )       
\end{equation}

where we have again introduced the IR length scale R.

\subsection{The beta function.}

As before, we make the identification $    \log \frac{R}{x_0}          \to      t        $, substitute the relations between $\phi_0$, $\gamma_0$, $\chi_0$ and F,G,H(\eqref{phi0}, \eqref{gamma0}, \eqref{chi0}) and the value of $\lambda_3$\eqref{lambda3}. Therefore we get the beta function

\begin{equation}
\beta_F          = -  (2 \delta F  -\frac{GH}{4}      )
\end{equation}

which matches all earlier results.

\section{Beta function for $\delta$.}

\subsection{Overview of the calculation.}

As mentioned before b multiplies X inside the cosine and therefore can be interpreted as the field strength renormalization. $b^2$ is close to 4. This is large compared to F which is close to zero. It was mentioned that the beta functions of the Sine-Gordon are a power series expansion in two parameters, $\delta=b^2/4 -1$ is the other appropriate small parameter in which the expansion can be carried out. From the perspective of the boundary theory the correction to this modified coupling $\delta$ comes from two cosines combining together to give $(\partial X)^2$, this is the anomalous dimension and gives this beta function. This has been calculated before in an earlier paper \cite{Oak:2017trw}. 

From the bulk theory the leading contribution to the beta function has to come from a vertex of the type $g_{\mu \nu} \partial ^{\mu}\phi \partial^{\nu} \phi $ so that we have a structure similar to the boundary calculation and we can see that the two $\phi$'s correct the $g_{\mu \nu}$ which is associated to the $(\partial X)^2$ term. Next we note that the boundary kinetic term involves only diagonal components and therefore we can attempt to model the graviton by a dilaton which takes into account only the diagonal degress of freedom and is a scalar, thus simplifying the problem enormously.

\begin{center}

\begin{tikzpicture}[scale=3]

\draw [vector] (3 ,0) -- (4,0);
\node at (3.9,0.1) {$\vec{x}_3$};
\node at (3.5 ,.2) {$\vec{b}_3$};

\draw  (3,0) circle (1);

\draw [fermionbar](3,0) node[below] {$\lambda_{\sigma}$}  -- (2.35,.8) ;
\node at (2.5,0.75) {$\vec{x}_2$};
\node at (2.7 ,.5) {$\vec{b}_2$};

\draw [fermionbar](3,0) -- (2.35,-.8);
\node at (2.5,-0.75) {$\vec{x}_1$};
\node at (2.7 ,-.5) {$\vec{b}_1$};

\node at (3,-1.4) { Fig. 3. } ;
\node at (3,-1.6) {Witten diagram for the graviton-scalar-scalar vertex }      ;
\node at (3,-1.8) { for the beta function of the field strength renormalization.};

\end{tikzpicture}

\end{center}

\subsection{Fixing the coupling of the graviton-scalar-scalar vertex in the bulk.}

To compute the graviton-scalar-scalar vertex we want to look at the fluctuation about AdS

\begin{equation}
g_{\mu \nu}=\bar{g}_{\mu \nu}+h_{\mu \nu}
\end{equation}

$\bar{g}_{\mu \nu}$ is AdS. We want to simplify this by modeling the graviton as a dilaton. Therefore,

\begin{equation}
g^{\mu \nu}=e^{-\lambda_{\sigma} \sigma } \bar{g}^{\mu \nu}=(1-{\lambda_{\sigma} \sigma }) \bar{g}^{\mu \nu}
 \end{equation}

Therefore the kinetic term $\hf g^{\mu \nu} \partial_{\mu}\phi \partial_{\nu}\phi$ becomes 

\begin{equation}
\hf g^{\mu \nu} \partial_{\mu}\phi \partial_{\nu}\phi=\hf (1-{\lambda_{\sigma} \sigma }) \bar{g}^{\mu \nu} \partial_{\mu}\phi \partial_{\nu}\phi
\end{equation}

We treat the dilaton as a massive scalar with $m_{\sigma}\to 0$, the full action therefore becomes,

\begin{align}
S_{bulk}=& \int d^3 x       \sqrt{g}      [     \hf  (\partial \phi)^2 + \hf (m_{\phi}\phi)^2        + \hf  (\partial \chi)^2 + \hf (m_{\chi}\chi)^2              \\                     \nonumber                                     
&+ \hf  (\partial \gamma)^2 + \hf (m_{\gamma}\gamma)^2 -        \lambda_3 \phi \gamma \chi           -\hf {\lambda_{\sigma} \sigma } \bar{g}^{\mu \nu} \partial_{\mu}\phi \partial_{\nu}\phi+\hf (\partial \sigma)^2 +\hf m_{\sigma}^2\sigma^2]
\end{align}

The kinetic term in the boundary action is modified

\begin{align}
S_{boundary}=&   \frac{1}{4\pi}    \int d^2 x    [  (   1  +     \xi_0 )   (\partial_{\mu} \vec{X}) .  (\partial^{\mu} \vec{X})         + m^2 \vec X .\vec X          +   \frac{F}{a(0)^2} \cos ~ (\vec{b}_1 . \vec{X} )     \\
& +   \frac{G}{a(0)^2} \cos ~ (\vec{b}_2 . \vec{X} )  +   \frac{H}{a(0)^2}       \cos ~ (\vec{b}_3 . \vec{X} )           ]  
\end{align}

$\xi_0$ is related to the bulk field $\sigma$ whose boundary value $\sigma_0$ is equal to $\xi_0$ upto normalization. We vary the action with respect to $\xi_0$ to compute various correlators.

\subsection{The 3-point correlator calculation.}

To compute the beta function we want to calculate  the generating functional for the three point function as before, but this time for the vertex $-\hf \lambda_{\sigma}\sigma \bar{g} ^{\mu \nu}  \partial _{\mu}\phi \partial_{\nu} \phi $. To do this we again start with

\begin{equation}
S_{\sigma 3}=    -\hf \lambda_{\sigma}                    \int d^{d+1} y \sqrt{g} ~      \bar{g}^{\mu \nu}        \partial_{\mu}   \phi(y)            ~ \partial_{\nu} \phi(y)   ~  \sigma(y)   
\end{equation}

We, as before, put in the expressions for the bulk to boundary propagators for all the fields and simplify. Details in Appendix \eqref{app beta delta}. We get,

\begin{align}
S_{\sigma 3}= &       -2\pi(1+\delta)    \lambda_{\sigma}   \sigma_0 \phi_0^2              \int                       d   \left( {  x_1^2     \over     R^2}    \right)         d    \left(    {x_3^2       \over R^2}    \right)                                \log \frac{R}{x_0}   
       \\                        \nonumber                      
&             \left[          \frac{\Gamma(\frac{\Delta_2+\Delta_3-\Delta_1}{2})\Gamma(\frac{\Delta_2+\Delta_3+\Delta_1-d}{2})}{       \Gamma(\Delta_2)        \Gamma(\Delta_3)      }                         - \frac{\Gamma(\frac{\Delta_2+\Delta_3-\Delta_1}{2})\Gamma(\frac{\Delta_2+\Delta_3+\Delta_1-d+2}{2})}{     \Gamma(\Delta_2)        \Gamma(\Delta_3+1)      }                \right]    
\end{align}

$S_{\sigma3}$ is non-zero offshell but vanishes onshell where the square bracket is zero. This is resolved when $\lambda_{\sigma}$ is fixed (\eqref{lambdasigma} and the comment thereafter).

Putting in all the relative normalizations(\eqref{phi0}, \eqref{sigma xi0})  and the value of $\lambda_{\sigma}$ and we get,

\begin{align}
S_{\sigma 3}=       {      - F^2 (1+\delta)       \over      8      }           \frac{\xi_0}{4\pi}          \int                                           d^2   \left( {  x_1     \over     R}    \right)         d    \left(    {x_3^2       \over R^2}    \right)                               \log \frac{R}{x_0}    
\end{align}

\subsection{The Beta function.}

The kinetic term in the boundary theory, whose correction we are computing is

\begin{equation}
\frac{   1   +    \xi_0}{4\pi}             \int       d^2 x    ~             (     \partial X(x)  ) ^2
\end{equation}

Comparing this with the expression for $S_{\sigma3}$ above and as before, making the identification $    \log \frac{R}{x_0}          \to      t        $,  we immediately see the beta function(the t-derivative) is,

\begin{equation}
\beta_{\delta}=   {      - F^2 (1+\delta)       \over      8      }  
\end{equation}

which     matches 5.2.46 in \cite{Oak:2017trw}. This computation is correct upto O$(F^2 \delta)$. There are higher order corrections to it. The computation for the boundary theory using the usual field theoretic approach has been done by Amit et al \cite{Amit:1979ab}. In their paper(Section 7) they have a detailed analysis of the higher order corrections. These can be easily obtained by an order by order double expansion in F and $\delta$. Therefore the next term would be O($F^3$).

\section{Summary and Conclusions}

In this paper the beta function of a generalized Sine-Gordon theory  has been calculated using a bulk holographic dual. The boundary theory is a free theory deformed by a term $F \cos b.X$. The anomalous dimension is proportional to F and goes to zero as $F \to 0$. The bulk theory is dual to a free field theory in the boundary for F=0. The bulk fields (in addition to the graviton) introduced correspond to the Cosine perturbation.  The calculations have been done both in position space as well as momentum space. The boundary calculation is also done and it is shown that the results agree.

To compute the beta functions, two and three point correlation functions were computed from the bulk. We are in fact constructing the bulk dual of the free theory and including just those bulk fields and interaction vertices that are necessary to reproduce correlators of some specific boundary operators (cos b.X) which would give us the beta functions. The correlators computed from the bulk match the boundary calculations upto normalization. These normalizations and the bulk interaction vertex coupling were further fixed by comparing the two and three point correlators on both sides, which is done in the appendix. The bulk theory was constructed in such a way as to reproduce the correlators in the boundary theory, therefore agreement between results was expected.

The main motivation for doing this calculation is to understand the results of \cite{Sathiapalan:2017frk} better when interactions are involved. There the main example used was the free scalar theory and in this situation interactions are between composite operators. In 1+1 dimension, the cosine is one of the most interesting example of such operators and besides being related to string theory,  has applications in 1+1 dimensional condensed matter systems, such as  the X-Y model \cite{Amit:1979ab}. 

The model is also motivated by the first quantized description of a string propagating in a tachyon background. The beta function gives the equation of motion for the tachyon. The model also has a wave function renormalization which results in a beta function for the string theory dilaton coupling. The boundary calculation in this paper uses techniques derived from the exact RG, as used in \cite{Oak:2017trw}  for the usual Sine-Gordon model.   The main idea for the bulk  calculation in position space is to identify the beta function with the coefficient of a logarithmic deviation from the canonical scaling of a two point function. This is based on the technique described in \cite{Sathiapalan:1987ic,polyakov} and is suitable for holographic computations. In the bulk momentum space calculation the  technique is to first solve the fourier transformed equations of motion order by order in the coupling of the bulk interaction vertex and then identify the log divergent terms in the solutions as described in \cite{Skenderis:2002wp}. All the results agree to the order calculated.

There are many further problems that need to be addressed. One technically interesting issue of course is to go to higher orders. This should constrain the bulk dual much more. The precise bulk dual of free scalar theory considered here, in particular the connection to higher spin theory in $AdS_3$ needs to be understood better. It would be interesting if one can say something about the IR fixed point of this theory by studying the bulk. One should remember that the underlying theory in the boundary is a free scalar theory. The interactions in the bulk  involve fields dual to composite operators. There are an infinite number of them - they can be identified with the momentum modes of the string theory tachyon. One expects that there should be a corresponding simple way to package these in the bulk also. This needs to be understood better. Finally, the ERG description of  composite operators and also  the map to a holographic theory in the presence of these operators, there are many complications. These can be studied in a controlled way in this model. We hope to return to these questions.

\section*{Future directions}
We would like to compute the aforementioned four point Witten diagrams and reproduce the beta functions for the Sine-Gordon model.

\section*{Acknowledgments}
P.O. would gratefully like to acknowledge many helpful discussions at the inital stage with Pinaki Banerjee.

\begin{appendices}

\section{Fixing relative normalization of the bulk and the boundary couplings and computing $\lambda_3$}

\subsection{Fixing $ \phi_0$, $\gamma_0$, $\chi_0$ and F, G, H relative  normalizations}

To compute the relative normalization between the bulk and the boundary for the couplings $\phi_0$ anf F we compare the generating functionals of the two point functions calculated for both sides.

The generating function for the two point function for the boundary theory is

\begin{equation}
GF_2=\frac{A_2}{4}      \frac{F^2}{(4\pi)^2}
\end{equation}

\begin{equation}
A_2=\int d^2x_1 d^2x_2 \frac{1}{(\vec{x}_1-\vec{x}_2)^{2\Delta}}
\end{equation}

The generating function for the two point function for the bulk is

\begin{equation}
S_2=\frac{2}{\pi}   A_2 \phi_0^2/(2!)
\end{equation}

Comparing $S_2$ and $GF_2$ we get

\begin{equation}\label{phi0}
\phi_0=\frac{1}{8\sqrt{\pi}}  F
\end{equation}

Similarly,

\begin{equation}\label{gamma0}
\gamma_0=\frac{1}{8\sqrt{\pi}}  G
\end{equation}

\begin{equation}\label{chi0}
\chi_0=\frac{1}{8\sqrt{\pi}}  H
\end{equation}

\subsection{Computing $\lambda_3$}

To compute $\lambda_3$ we compare the generating function for the three point function of the boundary theory and for the bulk theory.

The generating function for the three point function of the boundary theory is

\begin{equation}
GF_3=\frac{A_3}{4}   \frac{FGH}{(4\pi)^3}
\end{equation}

\begin{equation}
A_3=\int d^2x_1 d^2x_2  d^2x_3 \frac{1}{(\vec{x}_1-\vec{x}_2)^{\Delta_{123}}(\vec{x}_2-\vec{x}_3)^{\Delta_{231}}(\vec{x}_3-\vec{x}_1)^{\Delta_{312}}}
\end{equation}

Here $\Delta_{ijk}=\Delta_i  +  \Delta_j  -\Delta_k$.

For the bulk theory

\begin{equation}
S_3=-\frac{\lambda_3}{2\pi^2}        A_3    \phi_0      \gamma_0           \chi_0
\end{equation}

Comparing $GF_3$ and $S_3$ we get

\begin{equation}\label{lambda3}
\lambda_3= -4 (\pi)^{1/2}
\end{equation}

\subsection{Relative normalization between $\sigma_0$ and $\xi_0$}

To fix this we calculate the generating function of $\left\langle (\partial X(x_1))^2  (\partial X(x_2))^2  \right\rangle $ from the bulk and boundary and compare them.

Bulk:

\begin{equation}
GF_{\sigma_0^2}=\hf \sigma_0^2    \frac{\Gamma(\Delta+1)}{\pi^{d/2}\Gamma(\Delta-d/2)}   \int d^2x_1  d^2 x_2  \frac{1}{x_{12}^{2\Delta}} 
\end{equation}

\begin{equation}
GF_{\sigma_0^2}=  \frac {\sigma_0^2}{\pi}   A_2 
\end{equation}

Boundary:

\begin{equation}
GF_{\xi_0^2}=\frac{1}{2!}  \frac{\xi_0^2}{(4\pi)^2}     \int d^2 x_1   d^2 x_2         \left\langle (\partial X(x_1))^2  (\partial X(x_2))^2  \right\rangle  
\end{equation}

\begin{equation}
\left\langle (\partial X(x_1))^2  (\partial X(x_2))^2  \right\rangle=\frac{2}{x_{12}^4}
\end{equation}

Therefore

\begin{equation}
GF_{\xi_0^2}=\frac{1}{2!}  \frac{\xi_0^2}{(4\pi)^2}     \int d^2 x_1   d^2 x_2     \frac{2}{x_{12}^4}   
\end{equation}

\begin{equation}
GF_{\xi_0^2}=  \frac{\xi_0^2}{(4\pi)^2}             A_2     
\end{equation}

Comparing,

\begin{equation}\label{sigma xi0}
\sigma_0= \frac{\xi_0}{4 \sqrt{\pi}}
\end{equation}

\subsection{Fixing $\lambda_{\sigma}$ }

To fix $\lambda_{\sigma}$ we will compute $\left\langle  (\partial  X(x_1))^2 \cos bX(x_2)  \cos bX(x_3)    \right\rangle    $ from the bulk and the boundary and compare them.

Bulk:

From \eqref{GFsigma3}

\begin{equation}\label{1lamdasigma}
S_{\sigma3}=\frac{-\lambda_{\sigma}}{\pi^2} \sigma_0    \phi_0^2    (1+\delta)  A_3  [~]
\end{equation}

where $[~]=      \left[          \frac{\Gamma(\frac{\Delta_2+\Delta_3-\Delta_1}{2})\Gamma(\frac{\Delta_2+\Delta_3+\Delta_1-d}{2})}{       \Gamma(\Delta_2)        \Gamma(\Delta_3)      }                         - \frac{\Gamma(\frac{\Delta_2+\Delta_3-\Delta_1}{2})\Gamma(\frac{\Delta_2+\Delta_3+\Delta_1-d+2}{2})}{     \Gamma(\Delta_2)        \Gamma(\Delta_3+1)      }                \right]    $

Boundary:

We want to compute $\left\langle  (\partial  X(x_1))^2 \cos bX(x_2)  \cos bX(x_3)    \right\rangle   $

\begin{equation}\label{corr1}
\left\langle  (\partial  X(x_1))^2 \cos bX(x_2)  \cos bX(x_3)    \right\rangle_{non-vanishing}   =   2/4  \left\langle  (\partial  X(x_1))^2  \exp ibX(x_2)  \exp -i bX(x_3)    \right\rangle   
\end{equation}

We change to complex coordinates. $(\partial  X(x_1))^2      \to    4 \partial X_1 \bar{\partial}X_1$. We get,

\begin{equation}
2        \left\langle       \partial X_1 \bar{\partial}X_1  \exp ibX(x_2)  \exp -i bX(x_3)    \right\rangle 
\end{equation}

 We consider the product $ \exp i \alpha  \partial X_1     \exp i \beta   \bar{\partial}   X_1 $, where $\alpha$ and $\beta$ are close to zero.

\begin{align}
  \left\langle                      \exp i \alpha  \partial X_1     \exp i \beta   \bar{\partial}   X_1               \exp ibX(x_2)  \exp -i bX(x_3)    \right\rangle       \\     =         -\alpha \beta       \left\langle    \partial X_1 \bar{\partial}X_1  \exp ibX(x_2)  \exp -i bX(x_3)    \right\rangle 
\end{align}

is the part to linear order in $\alpha  \beta$. The coefficient of the $-\alpha \beta/2$ term will give us the correlator \eqref{corr1}.

Now,

\begin{align}
  \left\langle                      \exp i \alpha  \partial X_1     \exp i \beta   \bar{\partial}   X_1               \exp ibX(x_2)  \exp -i bX(x_3)    \right\rangle     
\end{align}

\begin{align}
=&  \exp          \bigg       [   \frac{1}{4}  \int d^2 z d^2 z'   \bigg (  \delta(z-z_1) \delta(\bar{z}-\bar{z_1})         \alpha      \partial       + \delta(z-z_1) \delta(\bar{z}-\bar{z_1})     \beta      \bar{\partial}         \\             \nonumber
&    + \delta(z-z_1) \delta(\bar{z}-\bar{z_1})   b    + \delta(z-z_1) \delta(\bar{z}-\bar{z_1})  (- b)        \bigg   )        \bigg   (   ln(z-z')(\bar{z}-\bar{z'} )    \bigg         )                 \\         \nonumber
&           \bigg (  \delta(z-z_1) \delta(\bar{z}-\bar{z_1})         \alpha      \partial       + \delta(z-z_1) \delta(\bar{z}-\bar{z_1})     \beta      \bar{\partial}         \\             \nonumber
&    + \delta(z'-z_1) \delta(\bar{z'}-\bar{z_1})   b    + \delta(z'-z_1) \delta(\bar{z'}-\bar{z_1})  (- b)        \bigg   )             \bigg     ]
\end{align}

For
$b^2=4(1+\delta) $ the expression becomes

\begin{equation}
(-\alpha  \beta/2)  \frac{-2 b^2}{4}    \frac{1}{z_{12}^2 z_{13}^2 z_{23}^{2}}
\end{equation}

Therefore, the generating function from the boundary theory is,

\begin{equation}\label{2lambdasigma}
GF_{\sigma3}=     \frac{1}{2!}      \frac{-2\xi_0 F^2 (1+\delta) }{(4 \pi)^3}   A_3
\end{equation}

Comparing \eqref{1lamdasigma} and \eqref{2lambdasigma},

\begin{equation}\label{lambdasigma}
\lambda_{\sigma}=\frac{4 \sqrt{\pi}}{    \left[          \frac{\Gamma(\frac{\Delta_2+\Delta_3-\Delta_1}{2})\Gamma(\frac{\Delta_2+\Delta_3+\Delta_1-d}{2})}{       \Gamma(\Delta_2)        \Gamma(\Delta_3)      }                         - \frac{\Gamma(\frac{\Delta_2+\Delta_3-\Delta_1}{2})\Gamma(\frac{\Delta_2+\Delta_3+\Delta_1-d+2}{2})}{     \Gamma(\Delta_2)        \Gamma(\Delta_3+1)      }                \right] }
\end{equation}

Thus the square brackets cancel out in $S_{\sigma3}$. The correlator remains finite on-shell.

\section{The sub-leading term for $\beta_F$ using ERG on the boundary}\label{app ERG sub leading beta F}

The action of the evolution operator is

\begin{align}
 \int d^2 x_1 d^2 x_2  F_{x_1 x_2 t}\frac{\delta^2}{\delta X_1 \delta X_2}    \bigg  [ { e^{ib_3X_3}  +e^{-ib_3X_3} \over 2}    \bigg ]       \bigg  [ { e^{ib_2X_4}  +e^{-ib_2X_4} \over 2}    \bigg ]
\end{align}

Here $X_i$ means $X(x_i)$. Keeping terms that contribute we get,

\begin{align}
 \int d^2 x_1 d^2 x_2  F_{x_1 x_2 t}\frac{\delta^2}{\delta X_1 \delta X_2}    \bigg  [ {   e^{-ib_3X_3-ib_2X_4}  +e^{ib_3X_3+ib_2X_4} \over 4}    \bigg ]
\end{align}

In the last expression the two terms that conserve momenta have been retained. We look at the action of the evolution operator on the first term. The second term gives an identical contribution.

\begin{align}
 \int d^2 x_1 d^2 x_2  F_{x_1 x_2 t}\frac{\delta^2}{\delta X_1 \delta X_2}    \bigg  [ e^{ib_3X_3+ib_2X_4}    \bigg ]
\end{align}

\begin{equation}
=(-b_3^2 F_{33t}-b_2.b_3 F_{34t}-b_2.b_3F_{34t}-b_2^2F_{44t})e^{i(b_3+b_2)X_4}
\end{equation}

Here $X_3$ has been taylor expanded and brought to $X_4$. Therefore,

\begin{align}\label{finalergontwocosines}
    \bigg  [ {   e^{ib_3X_3+ib_2X_4}  +e^{-ib_3X_3-ib_2X_4} \over 4}    \bigg ]
\end{align}

becomes

\begin{equation}\label{costaylor}
\hf \cos(b_2+b_3)X_4=\hf cos b_1X_4
\end{equation}

Substituting \eqref{finalergontwocosines} and \eqref{costaylor} in \eqref{evolu oper on two cosines} we get

\begin{align}
&     \frac{GH}{(4 \pi)^2}        \int    \frac{d^2x_1 d^2x_2}{a^4}     \exp[-\hf (-b_3^2-b_2^2) F_{11t}+b_2.b_3 F_{12t} ]            \\      \nonumber
& \hf    \cos b_1 X(x_2) 
\end{align}

where $F_{12t}=-\hf ln\frac{(x_1-x_2)^2+a(t)^2}{(x_1-x_2)^2+a(0)^2}$

now we relabel $x_2-x_1 \to y$ and $x_2 \to x$ we get

\begin{align}
&=    \frac{GH}{8}        \int       \frac{dy^2 }{a(t)^2}                            e^{4t-\frac{b_3^2+b_2^2}{2}t}       \bigg (  \frac{y^2+a(t)^2}{y^2+a(0)^2}         \bigg )         ^{ - b_2.b_3  \over 2 }            \frac{1}{(4 \pi)}        \int   \frac{d^2x }{a(t)^2}  \cos b_1 X(x)
\end{align}

$a(t)$ is the IR cutoff therefore we drop $y^2$ from the numerator and integrate.

\begin{align}
&=   \frac{GH}{8}    e^{4t-\frac{b_3^2+b_2^2}{2}t}               a(t) ^{- b_2.b_3 -2 }                    {   (a(t)^2+a(0)^2)     ^{ {  b_2.b_3  \over 2} +1}   -a(0)^{ 2( { b_2.b_3  \over 2      } +1  )  }      \over { {  b_2.b_3  \over 2      } +1}   }            \\      \nonumber
& \frac{1}{(4 \pi)}        \int   \frac{d^2x }{a(t)^2}  \cos b_1 X(x)
\end{align}

dropping $a(0)^2$ from the first term.

\begin{align}
&=   \frac{GH}{8}    e^{4t-\frac{b_3^2+b_2^2}{2}t}                                {  1   -         \bigg(         {a(t) \over a(0)}           \bigg   )         ^{- 2( {  b_2.b_3  \over 2      } +1)}       \over { {  b_2.b_3  \over 2      } +1}   }                 \frac{1}{(4 \pi)}        \int   \frac{d^2x }{a(t)^2}  \cos b_1 X(x)
\end{align}

For $b_2.b_3$ close to --2 and for $b_2^2=b_3^2=4$ we get,

\begin{align}
   \frac{GH}{4}      t     \frac{1}{(4 \pi)}        \int   \frac{d^2x }{a(t)^2}  \cos b_1 X(x)
\end{align}

\section{Position space calculation for $\beta_F$ from the bulk for the sub-leading term.}\label{app sub leading beta f}

\begin{align}
S_3&=           -     \lambda_3          \int d^dx_1  \int d^dx_2 \int d^dx_3  \int d^{d+1} y ~ y_0^{-(d+1)}      C_{\Delta_1}            C_{\Delta_2} C_{\Delta_3}  \\       \nonumber
&   ~\phi_0    ~   \gamma_0       ~\chi_0     {   y_0^{\Delta_1+\Delta_2 +\Delta_3}       \over                 (y_0^2+ (\vec{y}-\vec{x_1})^2)^{\Delta_1}      (y_0^2    + (\vec{y}-\vec{x_1})^2)^{\Delta_2}             (y_0^2   + (\vec{y}-\vec{x_2})^2)^{\Delta_3}                               }
\end{align}

 After Feynman parameterization we get,

\begin{align}
&=        -     \lambda_3         \int d^d x_1   \int d^dx_2 \int d^dx_3  \int d^{d+1} y ~    C_{\Delta_1}            C_{\Delta_2} C_{\Delta_3}   ~\phi_0        ~   \gamma_0       ~\chi_0     \\       \nonumber
& \int d \alpha_1 d \alpha_2  d \alpha_3          \alpha_1^{\Delta_1-1}  \alpha_2^{\Delta_2-1}  \alpha_3^{\Delta_3-1}    \delta(\Sigma_{i=1}^3   \alpha_i-1)                                                  {              \Gamma \left(  \Sigma_{i=1}^{3} \Delta_i  \right)                      \over                            \Pi_{i=1}^3      \Gamma  \left( \Delta_i \right)              }                                 \\          \nonumber
&     {   y_0^{{-(d+1)} +\Delta_1+\Delta_2 +\Delta_3}       \over       (      y_0^2         +          (\vec{y}-\Sigma_{i=1}^{n} \alpha_i \vec{x}_i ) ^2              +       \Sigma_{i<j=1}^3  \alpha_i   \alpha_j ( \vec{x}_i - \vec{x}_j            )^2         )^{\Delta_1+\Delta_2+\Delta_3}                }
\end{align}

We do the $y_0$ and $\vec{y}$ integrals

\begin{align}
S_3&=          -     \lambda_3           {             \pi^{d/2}          \Gamma \left(     {    \Sigma_{i=1}^{3} \Delta_i           \over       2       }             \right)                        \Gamma \left(     {    \Sigma_{i=1}^{3} \Delta_i    -d          \over       2                          }             \right)                          \over                   2                    \Pi_{i=1}^3      \Gamma  \left( \Delta_i \right)                       }                               \int d^d x_1   \int d^dx_2 \int d^dx_3   ~    C_{\Delta_1}            C_{\Delta_2} C_{\Delta_3}   ~\phi_0        ~   \gamma_0       ~\chi_0     \\       \nonumber
& \int d \alpha_1 d \alpha_2  d \alpha_3                      {         \alpha_1^{\Delta_1-1}  \alpha_2^{\Delta_2-1}  \alpha_3^{\Delta_3-1}    \delta(\Sigma_{i=1}^3   \alpha_i-1)                                                                                                \over       (      \Sigma_{i<j=1}^3  \alpha_i   \alpha_j ( \vec{x}_{ij}       )^2         )^{           {\Delta_1+\Delta_2+\Delta_3                \over          2            }                 }                }           
\end{align}

Now we transform from $\alpha_i$'s to $\beta_i$'s. $\alpha_i = \beta_1 \beta_i $ for  $i \geq 2$, $\alpha_1=\beta_1$. The Jacobian for n parameters is $\beta_1^{n-1}$. Here $n=3$.

\begin{align}
S_3&=           -     \lambda_3          {             \pi^{d/2}          \Gamma \left(     {    \Sigma_{i=1}^{3} \Delta_i           \over       2       }             \right)                        \Gamma \left(     {    \Sigma_{i=1}^{3} \Delta_i    -d          \over       2                          }             \right)                          \over                   2                    \Pi_{i=1}^3      \Gamma  \left( \Delta_i \right)                       }                               \int d^d x_1   \int d^dx_2 \int d^dx_3   ~    C_{\Delta_1}            C_{\Delta_2} C_{\Delta_3}   ~\phi_0        ~   \gamma_0       ~\chi_0     \\       \nonumber
&            \int d \beta_1 d \beta_2  d \beta_3              {     \beta_2^{\Delta_2-1}  \beta_3^{\Delta_3-1}  [\delta(\beta_1-1/(1+\beta_2+\beta_3))]/ (1+\beta_2+\beta_3)      \over   \beta_1               (          \beta_2x_{12}^2+\beta_3x_{13}^2+\beta_2\beta_3 x_{23}^2  )^{\Delta_1+\Delta_2+\Delta_3   \over 2}                }
\end{align}

After doing the $\beta_1$ integral we get

\begin{align}
S_3&=          -     \lambda_3           {             \pi^{d/2}          \Gamma \left(     {    \Sigma_{i=1}^{3} \Delta_i           \over       2       }             \right)                        \Gamma \left(     {    \Sigma_{i=1}^{3} \Delta_i    -d          \over       2                          }             \right)                          \over                   2                    \Pi_{i=1}^3      \Gamma  \left( \Delta_i \right)                       }                               \int d^d x_1   \int d^dx_2 \int d^dx_3   ~    C_{\Delta_1}            C_{\Delta_2} C_{\Delta_3}   ~\phi_0        ~   \gamma_0       ~\chi_0     \\       \nonumber
&            \int  d \beta_2  d \beta_3              {     \beta_2^{\Delta_2-1}  \beta_3^{\Delta_3-1}        \over             (     \beta_2x_{12}^2+\beta_3x_{13}^2+\beta_2\beta_3 x_{23}^2  )^{\Delta_1+\Delta_2+\Delta_3   \over 2}                }
\end{align}

\begin{align}
S_3&=          -     \lambda_3           {             \pi^{d/2}          \Gamma \left(     {    \Sigma_{i=1}^{3} \Delta_i           \over       2       }             \right)                        \Gamma \left(     {    \Sigma_{i=1}^{3} \Delta_i    -d          \over       2                          }             \right)                          \over                   2                    \Pi_{i=1}^3      \Gamma  \left( \Delta_i \right)                       }                               \int d^d x_1   \int d^dx_2 \int d^dx_3   ~    C_{\Delta_1}            C_{\Delta_2} C_{\Delta_3}   ~\phi_0        ~   \gamma_0       ~\chi_0     \\       \nonumber
&            \int  d \beta_2  d \beta_3              {     \beta_2^{\Delta_2-1}  \beta_3^{\Delta_3-1}        \over            (             \beta_2(x_{0}^2+x_{12}^2)+\beta_3(x_{0}^2+x_{13}^2)+\beta_2\beta_3(   x_{0}^2      +   x_{23}^2 )     )^{\Delta_1+\Delta_2+\Delta_3   \over 2}        }
\end{align}

Here we have introduced $x_0^2$'s in the denominator($x_0^2 \to 0$). These act as UV regulators. We do the $\beta_2$ and $\beta_3$ integrals. Any factors of $\delta$ coming from the two beta functions from the two beta integrals contribute at $O(\delta \phi_0 \gamma_0 \chi_0)$. Therefore they are dropped. We set $d=2$ and $\Delta_2=2(1+\delta)$, particle 2 is offshell. We substitue $C_{\Delta_i}$'s. We set $\vec{x}_1$ to zero using translation invariance, multiply and divide by $R$, the IR cut-off. Therefore, the integral simplifies to

\begin{align}
S_3=- {\lambda_3         \over       2        \pi^2         }        \int                         {              d^2 x_1   d^2 x_2  d^2 x_3           \over            R^6         }                     {          \phi_0 \gamma_0   \chi_0                   \over         \left(      { x_0^2+x_2^2     \over R^2    }        \right)     ^{(1+\delta)}     \left(      { x_0^2+x_{3}^2     \over R^2    }        \right) ^{(1-\delta)}    \left(      {   x_{0}^2    + x_{23}^2   \over R^2    }        \right) ^{(1+\delta)}       }
\end{align}

We will now calculate the log divergent term. The $x_2$ integral is,

\begin{align}\label{x2sqroriginal}
      \int            {      d^2   x_2              \over        R^2       }               {          1                 \over         \left(      { x_0^2+x_2^2     \over R^2    }        \right)     ^{(1+\delta)}     \left(      { x_0^2+x_{3}^2     \over R^2    }        \right) ^{(1-\delta)}    \left(      {   x_{0}^2    + x_{23}^2   \over R^2    }        \right) ^{(1+\delta)}       }
\end{align}

The log divergent contributions come from the two regions, when (i) $\vec{x}_2   \to             \vec{x}_3$, (ii) $\vec{x}_2 \to 0$.

(i)$\vec{x}_2   \to             \vec{x}_3$.

Set $\vec{y}=\vec{x}_2-\vec{x}_3$. At $\vec{x}_2= \vec{x}_3$, $\vec{y}=0$.

\begin{align}\label{x2sqr2}
        \int                          {          d^2   y    \over     R^2}               {                  1         \over         \left(      { x_0^2+(\vec{y}+\vec{x}_3 )^2    \over R^2    }        \right)     ^{(1+\delta)}     \left(      { x_0^2+x_{3}^2     \over R^2    }        \right) ^{(1-\delta)}    \left(      {   x_{0}^2    + y^2   \over R^2    }        \right) ^{(1+\delta)}       }
\end{align}

We taylor expand the first term in the denominator. We get,

\begin{align}
        \int                          {          d^2   y    \over     R^2}               {                                1    -(1+\delta)                  \left(         {       \vec{y}^2            -2 \vec{x}_3  . \vec{y}   \over  x_0^2+\vec{x}_3 ^2      }          \right)                        
                      \over         \left(      { x_0^2+\vec{x}_3 ^2            \over R^2    }        \right)     ^{(1+\delta)}     \left(      { x_0^2+x_{3}^2     \over R^2    }        \right) ^{(1-\delta)}    \left(      {   x_{0}^2    + y^2   \over R^2    }        \right) ^{(1+\delta)}       }                          
\end{align}

We drop the $\delta$ term. It is higher order. We look at,

\begin{align}
        \int                          {          d^2   y    \over     R^2}               {                                    -                  \left(         {      \vec{y}^2          
             -2 \vec{x}_3  . \vec{y}   \over                  x_0^2+\vec{x}_3 ^2           }          \right)        
                      \over         \left(      { x_0^2+\vec{x}_3 ^2            \over R^2    }        \right)         \left(      { x_0^2+x_{3}^2     \over R^2    }        \right)    \left(      {   x_{0}^2    + y^2   \over R^2    }        \right)       }               
\end{align}

Add and subtract $x_0^2$.

\begin{align}
        \int                          {          d^2   y    \over     R^2}               {                                    -                  \left(         {    x_0^2    +   \vec{y}^2          -x_0^2
             -2 \vec{x}_3  . \vec{y}   \over                  x_0^2+\vec{x}_3 ^2           }          \right)        
                      \over         \left(      { x_0^2+\vec{x}_3 ^2            \over R^2    }        \right)         \left(      { x_0^2+x_{3}^2     \over R^2    }        \right)    \left(      {   x_{0}^2    + y^2   \over R^2    }        \right)       }               
\end{align}

The $   x_0^2    +   \vec{y}^2  $ term cancels in the numerator and the denominator. We drop that.

\begin{align}
        \int                          {          d^2   y    \over     R^2}               {                                    -                  \left(         {           -x_0^2
             -2 \vec{x}_3  . \vec{y}   \over                  x_0^2+\vec{x}_3 ^2           }          \right)        
                      \over         \left(      { x_0^2+\vec{x}_3 ^2            \over R^2    }        \right)         \left(      { x_0^2+x_{3}^2     \over R^2    }        \right)    \left(      {   x_{0}^2    + y^2   \over R^2    }        \right)       }             
\end{align}

\begin{equation}
\int        d^2 y           (    -2 \vec{x}_3  . \vec{y}       )            =-2\int_0^{2 \pi} y dy d\theta_{3y}               x_3 y \cos \theta_{3y} =0
\end{equation}

Therefore we drop this term.  We get

\begin{align}
        \int                          {          d^2   y    \over     R^2}               {                                    -                  \left(         {           -x_0^2
          \over                  x_0^2+\vec{x}_3 ^2           }          \right)        
                      \over         \left(      { x_0^2+\vec{x}_3 ^2            \over R^2    }        \right)         \left(      { x_0^2+x_{3}^2     \over R^2    }        \right)    \left(      {   x_{0}^2    + y^2   \over R^2    }        \right)       }                 
\end{align}

which in the limit $x_0 \to 0$ goes to zero.

We look at

\begin{equation}
 \left(      {   x_{0}^2    + y^2   \over R^2    }        \right) ^{(1+\delta)}      =   \left(      {   x_{0}^2    + y^2   \over R^{2}    }        \right)             \left(      1      +     \delta  \log       \left(              {      x_{0}^2    + y^2                \over             R^2              }        \right)                             \right) 
\end{equation}

in the denominator. Again drop the $\delta$ term. Set $\vec{x}_3=\vec{R}$,

\begin{align}
        \int                          {          d^2   y    \over     R^2}               {                  1         \over         \left(      { x_0^2+R^2    \over R^2    }        \right)     ^{(1+\delta)}     \left(      { x_0^2+R^2     \over R^2    }        \right) ^{(1-\delta)}    \left(      {   x_{0}^2    + y^2   \over R^2    }        \right) ^{(1+\delta)}       }
\end{align}

The log divergent part is

\begin{align}
=   \pi        \int       ^{R^2}_{x_0^2}                   {          d   y^2    \over     R^2}               {                  1         \over          \left(      {   x_{0}^2    + y^2   \over R^2    }        \right) ^{(1+\delta)}       }
\end{align}

\begin{equation}
=     \pi      \log \frac{R^2}{x_0^2}   
\end{equation}

A similar computation for $\vec{x}_2 \to 0$ gives an identical contribution. The total contribution from both regions is,

\begin{equation}
=   2  \pi      \log \frac{R^2}{x_0^2}   
\end{equation}

The partition function becomes

\begin{align}
S_3=-       \hf           4    \lambda_3 \pi       \phi_0 \gamma_0   \chi_0              \int      d   \left( {  x_1^2     \over     R^2}    \right)         d    \left(    {x_3^2       \over R^2}    \right)                    \log \frac{R}{x_0}    
\end{align}

This expression corrects $b_1$ when $x_2 \to x_3$ and  $b_3$ when $x_2 \to x_1$ . These are both equal in magnitude. We only want the correction to $b_1$, therefore we divide the expression above by 2 to get the contribution of the generating functional to the beta function for $\cos b_1 X$.

\begin{align}
S_3=-         \lambda_3 \pi       \phi_0 \gamma_0   \chi_0              \int      d   \left( {  x_1^2     \over     R^2}    \right)         d    \left(    {x_3^2       \over R^2}    \right)                    \log \frac{R}{x_0}    
\end{align}

\section{Calculation for $\beta_{\delta}$.}\label{app beta delta}

We start with

\begin{equation}
S_{\sigma 3}=    -\hf \lambda_{\sigma}                    \int d^{d+1} y \sqrt{g} ~      \bar{g}^{\mu \nu}        \partial_{\mu}   \phi(y)            ~ \partial_{\nu} \phi(y)   ~  \sigma(y)   
\end{equation}

\begin{align}
= &   -\hf \lambda_{\sigma}   \sigma_0 \phi_0^2    C_{\Delta_1}        C_{\Delta_2} C_{\Delta_3}             \int d^{d+1} y \sqrt{g} ~            y_0^2                 ~                    \partial_{\mu}   \left(     {           y_0         \over          (y_0^2+(\vec{y}-\vec{x_1})^2)      }                   \right)^{\Delta_1}     \\                 \nonumber                     
&       ~ \partial^{\mu}               \left(     {           y_0         \over          (y_0^2+(\vec{y}-\vec{x_2})^2)      }                   \right)^{\Delta_2}     ~           \left(     {           y_0         \over          (y_0^2+(\vec{y}-\vec{x_3})^2)      }                   \right)^{\Delta_3}    
\end{align}

Set $\vec{x}_1=0$. Under inversion \cite{Freedman:1998tz},

\begin{equation}
\frac{y_0}{y_0^2+(\vec{x}-\vec{y})^2}    \to       ~~       \vec{x}'^2            \frac{y'_0}{y_0^{'2}+(\vec{x'}-\vec{y'})^2}      
\end{equation}

${\partial'}^{\mu}{y_0'}^{\Delta}={\partial'}^{0}{y_0'}^{\Delta}=\Delta {y_0'}^{\Delta-1}$.

\[{\partial'}^{\mu=0}{        \left(     {           y'_0         \over          ({y' _0}^2+(\vec{y'}-\vec{x'_i})^2)      }                   \right)}^{\Delta_i}=        \left(     {              \Delta_i         {y'_0}^{\Delta_i-1}       \over          ({y' _0}^2+(\vec{y'}-\vec{x'_i})^2) ^{\Delta_i}     }                   \right)-\left(     {              \Delta_i         {y'_0}^{\Delta_i}               ~2y_0              \over          ({y' _0}^2+(\vec{y'}-\vec{x'_i})^2) ^{\Delta_i+1}     }                   \right)
\]

\begin{align}
S_{\sigma 3}= &   -\hf \lambda_{\sigma}   \sigma_0 \phi_0^2    C_{\Delta_1}        C_{\Delta_2} C_{\Delta_3}                \Delta_1    \Delta_2                {x'}_2^{2\Delta_2}               {x'}_3^{2\Delta_3}                   \int d^{d+1} y'  ~            {y' _0}^{-(d+1)+2+ \Delta_1    -1   +\Delta_3+\Delta_2-1 }                   \\                 \nonumber                     
&        ~            \left(     {         1      \over          ({y' _0}^2+(\vec{y'}-\vec{x'_3})^2)      }                   \right)^{\Delta_3}      
                \left(     {               1      \over          ({y' _{0}}^2+(\vec{y'}-\vec{x'_2})^2) ^{\Delta_2}     }                   \right)
       \\                        \nonumber                
&         +     \hf \lambda_{\sigma}   \sigma_0 \phi_0^2    C_{\Delta_1}        C_{\Delta_2} C_{\Delta_3}                \Delta_1    \Delta_2                {x'}_2^{2\Delta_2}               {x'}_3^{2\Delta_3}                   \int d^{d+1} y'  ~            {y' _{0}}^{-(d+1)+2+ \Delta_1    -1  +\Delta_3+\Delta_2+1  }                   \\                 \nonumber                     
&        ~            \left(     {        1      \over          ({y' _{0}}^2+(\vec{y'}-\vec{x'_3})^2)      }                   \right)^{\Delta_3}      
                \left(     {            2                               \over          ({y' _{0}}^2+(\vec{y'}-\vec{x'_2})^2) ^{\Delta_2+1}     }                   \right)                
\end{align}

Thus, setting $\vec{x}_1$ to zero and using inversion we have reduced the number the of factors in the denominator from 3 to 2. This simplifies Feyman parameter integrals significantly. Now we Feynman parameterize and do $y'$ integrals. Doing the integrals and dropping pre-factors we get,

\begin{align}
 &    \frac{\pi}{2}   \frac{\Gamma(\frac{\Delta_2+\Delta_3-\Delta_1}{2})\Gamma(\frac{\Delta_2+\Delta_3+\Delta_1-d}{2})}{\Gamma(\Delta_3+\Delta_2)}         {\Gamma(\Delta_3+\Delta_2)         \over        \Gamma(\Delta_2)        \Gamma(\Delta_3)      }                 \\                 \nonumber                     
&           \int d \alpha_3       ~ d \alpha_2        ~      \delta(\alpha_3+\alpha_2-1)                  \alpha_3^{\Delta_3-1}         \alpha_2^{\Delta_2-1}                {      1             \over          \left(         \alpha_2          \alpha_3     \vec{x'}_{23}^2               \right)   ^{{\Delta_2+\Delta_3-\Delta_1                \over         2      }    }          }               
       \\                        \nonumber                
&       -2         \frac{\pi}{2}   \frac{\Gamma(\frac{\Delta_2+\Delta_3-\Delta_1}{2})\Gamma(\frac{\Delta_2+\Delta_3+\Delta_1-d+2}{2})}{\Gamma(\Delta_3+\Delta_2+1)}            {\Gamma(\Delta_3+\Delta_2+1)         \over        \Gamma(\Delta_2)        \Gamma(\Delta_3+1)      }             \\                 \nonumber                     
&        ~        \int d \alpha_3       ~ d \alpha_2        ~      \delta(\alpha_3+\alpha_2-1)                  \alpha_3^{\Delta_3}         \alpha_2^{\Delta_2-1}                                       {      1             \over          \left(           \alpha_2          \alpha_3     \vec{x'}_{23}^2               \right)   ^{{\Delta_2+\Delta_3-\Delta_1                \over         2      }    }          }     
\end{align}

The $\alpha$ integrals are:

The first,

\begin{align}
 \int d \alpha_3       ~ d \alpha_2        ~      \delta(\alpha_3+\alpha_2-1)                  \alpha_3^{\Delta_3-1}         \alpha_2^{\Delta_2-1}                {      1             \over          \left(        \alpha_2          \alpha_3                  \right)   ^{{\Delta_2+\Delta_3-\Delta_1                \over         2      }    }          }         =1     
\end{align}

The second,

\begin{align}
       \int d \alpha_3       ~ d \alpha_2        ~      \delta(\alpha_3+\alpha_2-1)                  \alpha^{\Delta_3}         \alpha^{\Delta_2-1}                                       {      1             \over          \left(                \alpha_2          \alpha_3             \right)   ^{{\Delta_2+\Delta_3-\Delta_1                \over         2      }    }          }         =1/2
\end{align}

 Use $(\vec{x'}^{2\Delta}=1/\vec{x}^{2\Delta})$ and
 
  $(\vec{x'}-\vec{y'})^{2(\Delta_1+\Delta_2-\Delta_3)/2}=(\vec{x}-\vec{y})^{2(\Delta_1+\Delta_2-\Delta_3)/2}/(\vec{x}^{2(\Delta_1+\Delta_2-\Delta_3)/2} \vec{y}^{2(\Delta_1+\Delta_2-\Delta_3)/2})$.

\begin{align}\label{GFsigma3}
S_{\sigma 3}= &       -\frac{\pi}{4}   \lambda_{\sigma}   \sigma_0 \phi_0^2    C_{\Delta_1}        C_{\Delta_2} C_{\Delta_3}                \Delta_1    \Delta_2        \int d^2x_1     d^2x_2          d^2x_3                         
       \\                        \nonumber                
 &                        {      1             \over          \left(               \vec{x}_{23}^2               \right)   ^{{\Delta_2+\Delta_3-\Delta_1                \over         2      }    }          }        {           1       \over           (x_2^2)^{ ^{{\Delta_2-\Delta_3+\Delta_1                \over         2      }    }     }       }                   {           1       \over           (x_3^2)^{ ^{{-\Delta_2+\Delta_3+\Delta_1                \over         2      }    }     }       }             
       \\                        \nonumber                
&             \left[          \frac{\Gamma(\frac{\Delta_2+\Delta_3-\Delta_1}{2})\Gamma(\frac{\Delta_2+\Delta_3+\Delta_1-d}{2})}{       \Gamma(\Delta_2)        \Gamma(\Delta_3)      }                         - \frac{\Gamma(\frac{\Delta_2+\Delta_3-\Delta_1}{2})\Gamma(\frac{\Delta_2+\Delta_3+\Delta_1-d+2}{2})}{     \Gamma(\Delta_2)        \Gamma(\Delta_3+1)      }                \right]    
\end{align}

Where now we have explicitly written integrals over the boundary coordinates(which were suppressed earlier). We insert a UV cutoff $x_0^2$ as before and multiply and divide by powers of $R^2$, we get

\begin{align}
S_{\sigma 3}= &       -\frac{\pi}{4}   \lambda_{\sigma}   \sigma_0 \phi_0^2    C_{\Delta_1}        C_{\Delta_2} C_{\Delta_3}                \Delta_1    \Delta_2                              \int                          d   \left( {  x_1^2     \over     R^2}    \right)                 d    \left(    {x_2^2       \over R^2}    \right)                   d    \left(    {x_3^2       \over R^2}    \right)                                                 
       \\                        \nonumber                
 &                        {      1             \over          \left(       {        x_0^2      +        \vec{x}_{23}^2            \over        R^2       }          \right)   ^{{\Delta_2+\Delta_3-\Delta_1                \over         2      }    }          }        {           1       \over           (   {        x_0^2      +        \vec{x}_{2}^2            \over        R^2       } )^{ ^{{\Delta_2-\Delta_3+\Delta_1                \over         2      }    }     }       }                   {           1       \over           (   {        x_0^2      +        \vec{x}_{3}^2            \over        R^2       } )^{ ^{{-\Delta_2+\Delta_3+\Delta_1                \over         2      }    }     }       }             
       \\                        \nonumber                
&             \left[          \frac{\Gamma(\frac{\Delta_2+\Delta_3-\Delta_1}{2})\Gamma(\frac{\Delta_2+\Delta_3+\Delta_1-d}{2})}{       \Gamma(\Delta_2)        \Gamma(\Delta_3)      }                         - \frac{\Gamma(\frac{\Delta_2+\Delta_3-\Delta_1}{2})\Gamma(\frac{\Delta_2+\Delta_3+\Delta_1-d+2}{2})}{     \Gamma(\Delta_2)        \Gamma(\Delta_3+1)      }                \right]    
\end{align}

$C_{\Delta_i}=1/\pi$. The square bracket vanishes. This gets renormalized when we fix $\lambda_{\sigma}$. We have taken particle 2 offshell. Therefore $ \Delta_2=  2(1+\delta)  $. $\Delta_1=2$. 

We get the same $x_2$ integral as before \eqref{x2sqroriginal}. The contribution from the $x_2$ integral from before is

\begin{equation}
   4  \pi      \log \frac{R}{x_0}   
\end{equation}

The contribution to the generating function is half of this as before.

\begin{equation}
   2  \pi      \log \frac{R}{x_0}   
\end{equation}

Therefore $S_{\sigma 3}$ becomes

\begin{align}
S_{\sigma 3}= &       -2\pi(1+\delta)    \lambda_{\sigma}   \sigma_0 \phi_0^2              \int                       d   \left( {  x_1^2     \over     R^2}    \right)         d    \left(    {x_3^2       \over R^2}    \right)                                \log \frac{R}{x_0}   
       \\                        \nonumber                      
&             \left[          \frac{\Gamma(\frac{\Delta_2+\Delta_3-\Delta_1}{2})\Gamma(\frac{\Delta_2+\Delta_3+\Delta_1-d}{2})}{       \Gamma(\Delta_2)        \Gamma(\Delta_3)      }                         - \frac{\Gamma(\frac{\Delta_2+\Delta_3-\Delta_1}{2})\Gamma(\frac{\Delta_2+\Delta_3+\Delta_1-d+2}{2})}{     \Gamma(\Delta_2)        \Gamma(\Delta_3+1)      }                \right]    
\end{align}

\end{appendices}

\end{document}